\documentclass{moriond}
\usepackage{lineno}

\def\Journal#1#2#3#4{{#1} {\bf #2}, #3 (#4)}

\def\NIMA{{\em Nucl. Instrum. Methods} A}

\def\PLB{{\em Phys. Lett.}  B}
\def\PRL{\em Phys. Rev. Lett.}

\def\be{\begin{equation}}
\def\ee{\end{equation}}
\def\bea{\begin{eqnarray}}
\def\eea{\end{eqnarray}}

\newcommand{\Photo}{}

\begin{document}
\vspace*{4cm}
\title{AN OVERVIEW OF STAR JET AND HIGH $p_{\mathrm{T}}$ RESULTS}

\author{SAEHANSEUL OH for the STAR Collaboration}

\address{Lawrence Berkeley National Laboratory,\\
1 Cyclotron Rd, Berkeley, CA 94720, USA }

\maketitle\abstracts{
These proceeding present recent measurements by the STAR Collaboration of jet production in $p$+$p$ and Au+Au collisions at $\sqrt{s_{\mathrm{NN}}} = 200$ GeV. We 
focus on jet yields, substructure, and heavy-flavor production, and their modification in the Quark-Gluon Plasma due to jet quenching.
}

\section{Jet measurements in STAR}
\label{sec:Intro}
The primary goal of the relativistic heavy ion physics program is to investigate the properties of Quark-Gluon Plasma (QGP), where the structure and dynamics of the system are governed by sub-hadronic degrees of freedom under extremely high energy density and temperature~\cite{HIjet}.
Jets are a well understood tool to explore the QGP properties, and interact with the QGP as they traverse it. 
Jet quenching in relativistic heavy ion collisions is characterized by three related phenomena: jet energy loss, modification of jet substructure, and medium-induced jet acoplanarity. 
Recent jet measurements in STAR cover various aspects of jet properties in $p$+$p$ collisions and jet quenching in heavy ion collisions, with the help of advances in measurement techniques, e.g.~HardCore and Matched jets~\cite{STARaj} and semi-inclusive jet measurements with event mixing~\cite{STARhj}. 
Such techniques precisely control the large fluctuating background in heavy ion collisions, enabling jet measurements in heavy ion collisions over a broad range in jet transverse momentum ($p_{\mathrm{T,jet}}$) and jet radius ($R$).
The STAR detector measures both charged constituents with the Time Projection Chamber (TPC), and neutral constituents with the Barrel Electromagnetic Calorimeter (BEMC) for jet reconstruction~\cite{starDet}.
Jets are nominally reconstructed with the anti-$k_{\mathrm{T}}$ algorithm with various $R$ from 0.2 to 0.6. 
Selected jet measurements from STAR are presented in these proceedings.

\section{Inclusive and semi-inclusive jet yields}
\label{sec:JetYield}
STAR has reported the first inclusive charged-particle jet yields in Au+Au collisions at $\sqrt{s_{\mathrm{NN}}} = 200$ GeV~\cite{STARij}, and released preliminary full jet yield measurements in the same collision system~\cite{STARfj}.
The charged-particle jet yield in central Au+Au collisions is observed to be suppressed compared to the binary-collision scaled yield in peripheral Au+Au collisions and that calculated by PYTHIA for $p$+$p$ collisions. 
The suppression is consistent with that observed for inclusive hadron yields at high $p_{\mathrm{T}}$~\cite{STARhRaa}.

Semi-inclusive jet measurements report jet yields in the recoil azimuthal region of trigger particles. 
The uncorrelated background in heavy ion collisions is precisely represented using mixed events, and is subtracted leaving only the yield of jets that are correlated with the trigger particles. 
Semi-inclusive charged-particle jet yields have been reported for trigger hadrons with $9 < p_{\mathrm{T}} < 30$ GeV/$c$ (h +jet)~\cite{STARhj}, and direct photon and $\pi^{0}$ triggers ($\gamma_{\mathrm{dir}}$+jet and $\pi^{0}$+jet, respectively)~\cite{STARgj}.
Recoil jets from trigger particles of different species may experience different path lengths in the medium, and originate from different parton flavors or energies. 
Thus, comparison of semi-inclusive jet measurements with different trigger particles may help to disentangle different elements of the jet quenching process. 
From the semi-inclusive charged-particle jet yields, similar levels of yield suppression between $\gamma_{\mathrm{dir}}$+jet and $\pi^{0}$+jet measurements in central Au+Au collisions are observed, requiring further theoretical advances in such channels of jet production and energy loss.

\begin{figure}
\centerline{\includegraphics[width=0.94\linewidth]{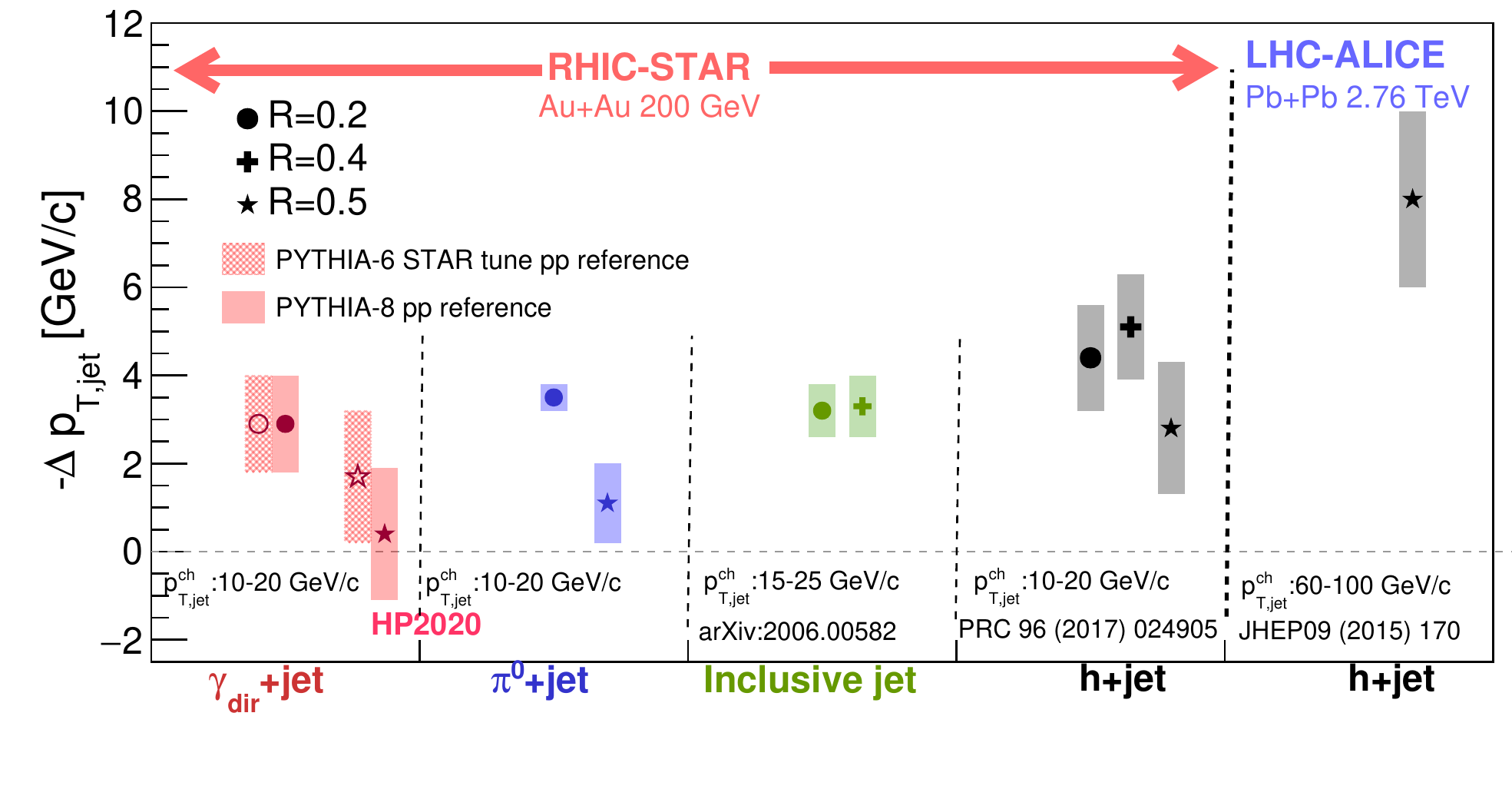}}
\caption[]{Tabulation of $p_{\mathrm{T,jet}}^{\mathrm{ch}}$ shift ($-\Delta p_{\mathrm{T,jet}}^{\mathrm{ch}}$) for $\gamma_{\mathrm{dir}}$+jet, $\pi^{0}$+jet, inclusive jet, h+jet measurements at RHIC, and h+jet at the LHC. Note that only charged-particle jet results are included, and $p_{\mathrm{T,jet}}^{\mathrm{ch}}$ ranges are different among measurements.}
\label{fig:jetptshift}
\end{figure}

Jet quenching is commonly quantified by measuring yield suppression at fixed $p_{\mathrm{T,jet}}$, such as $R_{\mathrm{AA}}$ and $I_{\mathrm{AA}}$, but this is only indirectly related to jet energy loss through the shape of the spectrum. 
We therefore parameterize the energy loss by a $p_{\mathrm{T,jet}}$-shift ($-\Delta p_{\mathrm{T,jet}}$) between spectra in central heavy ion collisions and their $p$+$p$ references, so that the effect from the shape of the spectrum is removed. 
Figure~\ref{fig:jetptshift} shows the results of $-\Delta p_{\mathrm{T,jet}}$ for different measurement channels, including inclusive and semi-inclusive jet measurements by STAR and the hadron+jet measurement by ALICE at the LHC. 
While the level of energy loss is consistent among STAR measurements, the ALICE hadron+jet result for a higher $p_{\mathrm{T,jet}}$ range indicates larger energy loss at the LHC energy.

\section{Jet substructure}
\label{sec:JetStructure}
Modification of jet substructure in heavy ion collisions has been studied in STAR with jet fragmentation functions and jet shapes~\cite{STARFF}.
The preliminary jet fragmentation functions in 40-60\% peripheral Au+Au collisions are based on a semi-inclusive approach and charged-particle jets. 
They are consistent with the PYTHIA-8 estimation for $p$+$p$ collisions within $15 < p_{\mathrm{T,jet}}^{\mathrm{ch}} < 30$ GeV/$c$. 
This measurement will be extended to the most central Au+Au collisions in the forthcoming publication. 
Jet shapes, which provide information about the radial distribution of the momentum carried by jet constituents, are measured with HardCore full jets, i.e.~constituent tracks and towers with $p_{\mathrm{T}} > 2.0$ GeV/$c$ and $E_{\mathrm{T}} > 2.0$ GeV are used for jet reconstruction. 
Such a jet selection provides an effective rejection of combinatorial jets, although it introductions a selection bias.
The preliminary jet shapes in STAR are observed to be broader than those at the LHC energies, with the $p_{\mathrm{T,jet}}$ ranges lower at RHIC than the LHC. 
Additionally, the event-plane dependence of jet shapes is investigated by classifying jets based on their azimuthal angle with respect to the second-order event plane. 
The jet shape function is enhanced for out-of-plane jets in comparison to in-plane jets, particularly at larger distances from the jet axis and lower $p_{\mathrm{T}}$ tracks.
This may indicate larger jet quenching effect for out-of-plane jets relative to in-plane jets due to different path-lengths.

Extensive jet substructure measurements in $p$+$p$ collisions have been published, and these will be the baseline for the corresponding measurements in $p$+A and heavy ion collisions. 
Such measurements include the shared momentum fraction ($z_{\mathrm{g}}$) and the groomed jet radius ($R_{\mathrm{g}}$) via the SoftDrop algorithm~\cite{STARzg}, and jet mass and groomed jet mass~\cite{STARm}. 
These results are compared to leading order Monte Carlo generators, such as PYTHIA-6, PYTHIA-8, and HERWIG-7. 
While the RHIC-tuned PYTHIA-6 quantitatively reproduces the measured data on these observables, PYTHIA-8 and HERWIG-7, which are tuned at the LHC energy,  do not agree with the data. 
The results enable further parameter tuning for these event generators at RHIC energies.

\section{Heavy flavor in jets}
\label{sec:HFjet}
\begin{figure}
\centerline{\includegraphics[width=0.52\linewidth]{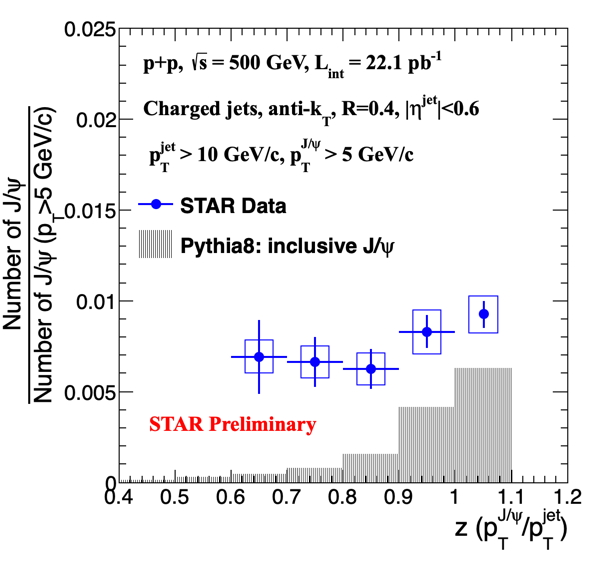}}
\caption[]{The normalized $z$ distributions for incluisve $J/\psi$ mesons produced within a jet compared to the PYTHIA 8 prediction. Data are normalized by the $J/\psi$ cross-section with $p_{\mathrm{T}}^{J/\psi} > 5$ GeV/$c$ at the $\sqrt{s} = 500$ GeV.}
\label{fig:JpsiJet}
\end{figure}

Heavy-flavor quarks are another channel of hard probes in relativistic heavy ion collisions, as they are produced at the early stage of collisions and their production can be well calculated with perturbative QCD.
In addition to direct measurements of open heavy flavor and quarkonium production yields, STAR has reported $J/\psi$ production in jets, via the transverse momentum fraction of $J/\psi$ meson with respect to the charged-particle jet, i.e.~$z(J/\psi) \equiv p_{\mathrm{T,J/\psi}}/p_{\mathrm{T,jet}}^{\mathrm{ch}}$, in $p$+$p$ collisions at $\sqrt{s} = 500$ GeV. 
Such measurements are motivated by their capability to differentiate various $J/\psi$ production models.
Figure~\ref{fig:JpsiJet} shows the preliminary ratio of $J/\psi$ within a jet to inclusive $J/\psi$, which is calculated by the number of $J/\psi$ for $p_{\mathrm{T},J/\psi} > 5$ GeV/$c$ and $p_{\mathrm{T,jet}}^{\mathrm{ch}} > 10$ GeV/$c$ to the total number of $J/\psi$ with $p_{\mathrm{T},J/\psi} > 5$ GeV/$c$.
The result shows discrepancy with the PYTHIA-8 estimation, including 1) more $J/\psi$ yields in jets in data, and 2) more isolated $J/\psi$ production in PYTHIA-8.  

In Au+Au collisions, $D^{0}$-meson, one of the open-charm hadrons, is used for the study of charm-quark interaction with the medium. 
Two-particle angular correlations between  $D^{0}$-meson and charged hadrons have been reported, where the $D^{0}$-meson serves as a proxy for charm-quark jets~\cite{STARD0h}. 
The 2-dimensional angular correlations in $(\phi, \eta)$ space reveal similar structure and centrality dependence to those from light-flavor correlations, indicating that the effective strength and centrality dependence of charm quark interactions with the hot and dense QCD medium are similar to those observed for light flavor. 
These results are complementary to previous studies, including $D^{0}$-meson spectra, $R_{\mathrm{AA}}$, and $v_{2}$, and provide a consistent picture for charm-quark measurements.

\section{Outlook}
\label{sec:Outlook}
Although only limited fraction of jet measurements from STAR is summarized in these proceedings, a wide variety of measurements are actively being performed in STAR. 
Several collision systems have been particularly utilized for jet measurements, including $p$+$p$, $p$+Au, Au+Au, Zr+Zr and Ru+Ru, and the system size dependence of jet quenching is additionally being investigated. 
These results at RHIC energies are complementary to those from the LHC, and require theoretical inputs to understand the data in the wide range of collision energies and system sizes. 
The upgrade of the STAR detector and the operating plan through the year of 2025 promise increased statistics and improved instrumentation, which will produce impactful physics results in the future.

\section*{Acknowledgments}
This work was supported by the Director, Office of Science, Office of Basic Energy Sciences, of the U.S. Department of Energy under Contract No. DE-AC02-05CH11231

\end{document}